\documentstyle[11pt]{article}
\begin{document}

\begin{titlepage}

\noindent{\today \hfill USITP 98-05}

\bigskip

\begin{center}
{\Large FROM TOPOLOGICAL TO 

\medskip

PARAMETRIZED FIELD THEORY}\\
\vspace{15mm}

{\large Nuno Barros e S\'{a}}\footnote{Email address: 
nunosa@vanosf.physto.se. Supported by grant PRODEP-Ac\c c\~ao 5.2.}

\vspace{3mm}

{\large Ingemar Bengtsson}\footnote{Email address: 
ingemar@vana.physto.se. Supported by NFR-project 07923.}
\vspace{5mm}

{\it $^{1,2}$Fysikum, Stockholms Universitet, Box 6730,

113 85 Stockholm, Sverige

\medskip
$^1$DCTD, Universidade dos A\c cores, R. M\~ae de Deus,

9500 Ponta Delgada, A\c cores, Portugal}
\vspace{15mm}

{\bf Abstract}\\
\ \\

\end{center}
It has been proposed to study the theory resulting from 
setting the gravitational constant to zero in the first 
order formalism for general relativity. In this letter we 
investigate this theory in the presence of matter fields, 
establish its equivalence with parametrized field theory 
on a flat background, and relate it to previous results 
in topological field theory (BF theory).

\ \\
\end{titlepage}
\newpage

\noindent Over the past ten years a bewildering array of 
diffeomorphism invariant field theories has been studied. 
In many cases they contain a finite number of degrees of 
freedom only, and they carry topological information about 
the manifold on which they are defined. By contrast, the most 
interesting theory of this kind remains that of Einstein, in 
which the metric occurs as a dynamical variable carrying two 
degrees of freedom per spatial point. Einstein's theory in 
2+1 dimensions occupies the middle ground. It can be 
formulated like the 3+1 dimensional theory, but 
it can also be formulated as a topological gauge theory in 
which the metric---or more precisely the metrical triad---occurs 
as a gauge field \cite{Achucarro}. When matter couplings are 
included the first formulation turns out to be superior 
\cite{Blencowe}. One is left wondering 
about the conditions under which a set of variables in a 
diffeomorphism invariant theory can be meaningfully identified 
as a spacetime metric. 

We will not try to answer the general question in this letter. 
Instead, we will point out that there is a set of ideas that can 
be used to provide a 3+1 dimensional illustration of the issues 
involved. First we will show that the Einstein-Hilbert action 
can be "short circuited" in a certain way so that the metric 
carries no degrees of freedom (as first observed by Smolin 
\cite{Smolin}). In effect we obtain a diffeomorphism invariant 
theory where the only solution is Minkowski space---also when 
matter fields are included. We then show that this model is 
closely related to the BF topological field theory defined by 
Horowitz \cite{Horowitz}. As in 2+1 gravity it is possible 
to reorganize the constraints of the model so that we obtain 
the constraint algebra familiar from general relativity. Our 
fourth and final point is that we can solve some of the 
constraints of our theory by means of a canonical 
transformation to gauge invariant variables. When 
this is done we recover precisely the parametrized field 
theories studied by Dirac and Kucha\v r \cite{Dirac}. 

We will assume that the metric that we define is non-degenerate. 
While rather foreign to topological field theory this assumption 
is natural in metrical theories. The same assumption has to be 
made in 2+1 gravity in order to show that the two formulations 
referred to above are indeed equivalent. In this as well as 
in our case there is a subtlety involved, and we will comment 
on this at the appropriate point. 

Our starting point is the first order action for gravity 
built from tetrads $e_I$ and connections ${\omega}_{IJ}$, 

\begin{equation} S_E = \frac{1}{8}\int {\epsilon}^{{\alpha}
{\beta}{\gamma}{\delta}}{\epsilon}^{IJKL}e_{{\alpha}I}
e_{{\beta}J}R_{{\gamma}{\delta}KL} + S_m \ . 
\end{equation}

\noindent Here $I, J, ...$ are internal indices that can 
be raised and lowered with a Minkowski metric ${\eta}_{IJ}$, 
the curvature tensor is 

\begin{equation} R_{{\alpha}{\beta}IJ} = \partial_{\alpha}
{\omega}_{{\beta}IJ} - \partial_{\beta}{\omega}_{{\alpha}IJ} + 
G{\omega}_{{\alpha}I}^{\ \ \ K}{\omega}_{{\beta}KJ} \ , \end{equation}

\noindent $G$ is the gravitational coupling constant, and 
this form of the first order action may differ from that of other references
by a redefinition $\omega_{IJ}\to G\omega_{IJ}$. $S_m$ is an action 
for matter fields that is independent of the connection and 
depends on the tetrad only through the metric tensor 

\begin{equation} g_{{\alpha}{\beta}} = e_{{\alpha}I}
e_{{\beta}J}{\eta}^{IJ} 
\ . \end{equation}

\noindent For definiteness we may choose an action for a 
scalar field, 

\begin{equation} S_m = - \frac{1}{2}\int \sqrt{-g}(
g^{{\alpha}{\beta}}\partial_{\alpha}{\varphi}\partial_{\beta}
{\varphi} + m^2{\varphi}^2) \ . \end{equation}

\noindent Our results will however be general and will not 
depend on this particular choice of the matter action.

The model that we will consider is obtained by setting $G = 0$ 
in the above action; 

\begin{equation} S = \frac{1}{4}\int {\epsilon}^{{\alpha}
{\beta}{\gamma}{\delta}}{\epsilon}^{IJKL}e_{{\alpha}I}
e_{{\beta}J}\partial_{\gamma}{\omega}_{{\delta}KL} + S_m \ . 
\end{equation}

\noindent This is a drastic operation. Instead of an $SO(3,1)$ 
connection we now have a collection of six $U(1)$ connections, 
and the tetrad is a collection of four gauge invariant vector 
fields. Moreover we will see that this theory does not have 
any local degrees of freedom (in the absence of matter fields). 
The use of the matrix ${\eta}_{IJ}$ to build a spacetime metric 
may therefore seem completely ad hoc. However, we will see 
that---at least when matter fields are included---it is 
actually quite well motivated. We do need ${\eta}_{IJ}$ to 
build the matter action $S_m$. 

The $G = 0$ "limit" was first studied by Smolin \cite{Smolin} 
in the context of Ashtekar's variables. Then the starting 
point is the self-dual form of the action, and the detailed 
results are quite different from ours since the equivalence 
between the two forms of the action breaks down when $G = 0$. 
In particular the model studied by Smolin has the same number 
of degrees of freedom as Einstein's theory (either Euclidean or
complex Lorentzian). 

Varying our action with respect to the connection yields 
an equation for the tetrad:

\begin{equation} {\epsilon}^{{\alpha}{\beta}{\gamma}{\delta}}
{\epsilon}^{IJKL}e_{{\gamma}K}e_{{\delta}L} = 0 \hspace{5mm} 
\Leftrightarrow \hspace{5mm} \partial_{\alpha}e_{{\beta}I} 
- \partial_{\beta}e_{{\alpha}I} = 0 \ . \end{equation} 

\noindent (Varying Einstein's action would yield an 
equation for the connection at this point---our action is 
indeed "short circuited".) This means that the tetrads 
are closed forms and that the only solution is (locally) Minkowski 
spacetime:

\begin{equation} e_{{\alpha}I} = \partial_{\alpha}f_I 
\hspace{5mm} \Rightarrow \hspace{5mm} g_{{\alpha}{\beta}} 
= \partial_{\alpha}f^I{\eta}_{IJ}\partial_{\beta}f^J \ . 
\end{equation}

\noindent Varying the tetrads one obtains a set of equations 
relating the connections to the tetrads and to the matter 
fields. These may be solved for the connection---they do 
not constrain either the tetrads or the matter fields. 
Finally, varying the matter fields leads to the usual 
field equations for matter propagating in a flat background. 
Hence we have a diffeomorphism invariant theory in which 
matter does not curve geometry. 

One may wonder whether there is anything special about 
flat space here? From the present point of view there is; 
although one may linearize the first order action around 
any connection ${\omega}^{(0)}_{IJ}$ that solves 
Einstein's equations and then proceed as above diffeomorphism 
invariance would be lost in the process. Unless 
${\omega}^{(0)}_{IJ} = 0$ one ends up with an action 
that contains fixed functions of the coordinates.

We will now analyze our model in more detail using the 
Hamiltonian formalism. Until further notice we set 
$S_m = 0$, that is to say that we do not include the matter 
degrees of freedom. We start by defining a new set 
of variables 

\begin{equation} B_{{\alpha}{\beta}}^I = - \frac{1}{2}
{\epsilon}^{IJKL}e_{[{\alpha}J}{\omega}_{{\beta}]KL} 
\ . \end{equation}

\noindent Provided that the tetrad is indeed non-degenerate 
this is a one-to-one transformation from the connections 
to the two forms $B^I$, having the inverse 

\begin{equation} \omega_{\alpha IJ}=\frac{1}{2}\epsilon_{IJKL}
e^{\beta K}\left( {B_{\alpha\beta}}^L-\frac{1}{2}
e^{\gamma L} e_{\alpha M} {B_{\beta\gamma}}^M\right) \ .
\end{equation} 

\noindent Hence we can perform this change of variables 
in the action. After a partial integration it becomes 

\begin{equation} S = \int {\epsilon}^{{\alpha}{\beta}
{\gamma}{\delta}}B_{{\alpha}{\beta}}^I\partial_{\gamma}
e_{{\delta}I} \ . \end{equation}

\noindent In this form the action is just four copies of the BF 
topological field theory studied by Horowitz \cite{Horowitz}. 
Hence we have shown that the $G = 0$ version of the Einstein 
action is equivalent to that sector of the abelian BF 
theory in which the tetrad is invertible.

The action has a large gauge invariance (larger than that 
of the Einstein action), namely 

\begin{equation} e_{{\alpha}I} \rightarrow e_{{\alpha}I} - 
\partial_{\alpha}{\Lambda}_I \  \end{equation}

\begin{equation} B_{{\alpha}{\beta}}^I \rightarrow B_{{\alpha}{\beta}}^I 
- \partial_{[{\alpha}}{\Lambda}_{{\beta}]}^I \ . \end{equation}

\noindent It follows that a non-degenerate tetrad can always be 
transformed so that it vanishes at any chosen point. The 
same difficulty occurs when one attempts to show the 
equivalence between the two possible formulations of 
2+1 gravity referred to above. It is a moderately 
embarrassing difficulty; a possible attitude to take 
in both cases is that the true starting point is a 
suitable form of the phase space action, in which this problem 
does not appear. Therefore we proceed with the calculation.

It is straightforward to perform a 3+1 decomposition of 
the action in the form that we arrived at. We get

\begin{equation} S = \int \dot{e}_{{\alpha}I}{\pi}^{aI} 
+ e_{tI}\partial_a{\pi}^{aI} - {\lambda}_a^I{\epsilon}^{abc}
\partial_be_{bI} \ , \end{equation}

\noindent where we renamed the components of the 
two form according to 

\begin{equation} {\pi}^{aI} = {\epsilon}^{abc}B_{bc}^I 
\hspace{1cm} {\lambda}_a^I = - 2B_{ta}^I \ . \end{equation}

\noindent Excluding the matter action there are 16 first class 
constraints of which 12 are independent, and there are 12 canonical 
variables per spatial point. Hence the model is devoid 
of dynamical degrees of freedom. The constraint algebra 
is abelian, which means that diffeomorphism invariance---which 
should be realized as a gauge freedom---is not manifest. 
This form of the phase space action is analogous to 
Witten's form of the phase space action for 2+1 gravity 
\cite{Achucarro}. Non-degeneracy of the tetrad is not 
a gauge invariant property in this formulation, and 
inclusion of the matter fields in the constraints is problematic. 

These problems can be avoided by a redefinition of the 
Lagrange multipliers. We wish to interpret the tetrad as 
giving rise to a metric, and we would therefore like to 
introduce the lapse function $N$ and the shift vector 
$N^a$ as multipliers. Any spacetime metric can be split 
into the spatial metric 

\begin{equation} q_{ab} = e_{aI}e_{bJ}{\eta}^{IJ} \end{equation} 

\noindent induced on a spatial hypersurface at constant $t$, 
together with the lapse and shifts 

\begin{equation} N = \frac{1}{\sqrt{-qg^{tt}}} \hspace{1cm} 
N^a = - \frac{g^{ta}}{g^{tt}} \ . \end{equation}

\noindent (Here $q$ is the determinant of the spatial metric, 
and we use a lapse function that is 
a tensor density of weight minus one.) These equations can be 
inverted so that the time component of the tetrad becomes 

\begin{equation} e_{tI} = \frac{N}{6}{\epsilon_I}^{JKL}
{\epsilon}^{abc}e_{aJ}e_{bK}e_{cL} + N^ae_{aI} \ . \label{17} 
\end{equation}

\noindent Our action does not single out a preferred metric, but 
if we trade the Lagrange multiplier $e_{tI}$ for the lapse and 
shifts it will acquire one, having a signature determined by 
the signature of the matrix ${\eta}_{IJ}$.

When we use this result in the phase space action we obtain 

\begin{equation} S = \int \dot{e}_{aI}{\pi}^{aI} - N{\cal H} 
- N^a{\cal H} - {\lambda}^I{\epsilon}^{abc}\partial_be_{cI} \ 
. \end{equation}

\noindent The constraints are the Hamiltonian and vector constraints, 
together with the constraints ${\phi}^a_I = 0$ already encountered:

\begin{equation} {\cal H} = \frac{1}{3!}{\epsilon^{IJK}}_L
{\epsilon}^{abc}e_{aI}e_{bJ}e_{cK}\partial_d{\pi}^{dL} \end{equation}

\begin{equation} {\cal H}_a = - e_{aI}\partial_b{\pi}^{bI} \end{equation}

\begin{equation} {\phi}^a_I = {\epsilon}^{abc}\partial_be_{cI}\ . 
\end{equation}

\noindent In this formulation the spatial diffeomorphisms are manifest, 
or almost so. In fact there is a simple combination of the constraints 
that generates spatial diffeomorphisms; 

\begin{equation} {\cal D}_a = {\cal H}_a + {\epsilon}_{abc}
{\pi}^{bI}{\phi}^c_I \ . \end{equation}

\noindent The constraint algebra has the following non-zero brackets: 

\begin{equation} \{{\cal H}[N], {\cal H}[M]\} = {\cal H}_a[(N\partial_aM 
- M\partial_aN)qq^{ab}] \end{equation}

\begin{equation} \{{\cal H}_a[N^a], {\cal H}[M]\} = {\cal H}
[{\cal L}_{\bar{N}}M] - {\phi}^a_I[{\epsilon^{IJK}}_L M N^b e_{bJ}
e_{aK} \partial_c{\pi}^{cL}] \ ,\end{equation}

\begin{equation} \{{\cal H}_a[N^a], {\cal H}_b[M^b]\} = {\cal H}_a
[{\cal L}_{\bar{N}}M^a] - {\phi}^a_I[{\epsilon}_{abc}N^bM^c
\partial_d{\pi}^{dI}] \ , \end{equation}

\noindent where ${\cal L}$ denotes the Lie derivative and the square bracket
denote smearing with test functions. On the constraint 
surface ${\phi}^a_I = 0$ this is the usual constraint algebra of general 
relativity. It is the fingerprint of diffeomorphism invariance in a metric
space in the Hamiltonian formulation, having the geometrical
interpretation \cite{Hojman} as the algebra of deformations of spatial 
hypersurfaces in a Lorentzian spacetime. From this point of view the
matrix ${\eta}_{IJ}$ 
is an object that is inserted in the phase space action precisely 
in order to make a geometrical interpretation of the solutions 
possible.

We may now adopt this first order action as a precise definition 
of our model. This is analogous to the ADM formulation of 2+1 gravity, 
and it has the double advantages that non-degeneracy of the metric 
can be consistently imposed, and that inclusion of matter degrees 
of freedom is straightforward. The latter will affect the form 
of the Hamiltonian and vector constraints, but they will not affect 
the constraint algebra. This follows from the assumption that 
only metrical couplings of the matter fields will be considered
(that is the case for bosonic fields and also for the standard coupling
of fermionic fields when setting $G=0$), 
together with the crucial bracket

\begin{equation} \{q_{ab}, {\cal H}[N]\} = N\frac{1}{2}e_{aI}
{\epsilon}^{IJKL}{\epsilon}^{cde}\partial_b(e_{cJ}e_{dK}e_{eL}) 
\ + \ {\small (a \leftrightarrow b)} \ . \end{equation}

\noindent The point here is the absence of any derivatives 
acting on the lapse function. (For a full explanation of this point 
as well as of all other properties of our constraint algebra, see 
Hojman et al. \cite{Hojman}.)

We now have a diffeomorphism invariant theory that describes 
matter propagating on a flat background. There are no local degrees 
of freedom attached to the geometry. One might expect that there 
should be a relation to parametrized field theory \cite{Dirac}, which 
achieves the same goal at the expense of introducing special 
"embedding variables" into the action. This is in fact the case. To 
see this we observe that, restricting ourselves to simply connected space,
we can solve some of our constraints explicitly: 

\begin{equation} {\phi}^a_I = {\epsilon}^{abc}\partial_be_{cI} = 0 
\hspace{.5cm} \Leftrightarrow \hspace{.5cm} e_{aI} = \partial_aX_I \ .
\end{equation}

\noindent We can now use Hamilton-Jacobi theory to effect a 
(singular) canonical transformation from the tetrad and its momentum 
to a new set of canonical pairs $(X_I, P^J)$ that coordinatize the 
constraint surface ${\phi}^a_I = 0$ modulo the gauge transformations 
generated by this constraint \cite{Newman}. To this end we choose a generating 
functional that depends on the "old" momenta and the "new" coordinates, 

\begin{equation} S_{pQ} = - \int \partial_aX_I{\pi}^{aI} \ . \end{equation}

\noindent Then the canonical transformation is given by 

\begin{equation} e_{aI} = - \frac{{\delta}S_{pQ}}{{\delta}{\pi}^a_I} 
= \partial_aX_I \end{equation}

\begin{equation} P^I = - \frac{{\delta}S_{pQ}}{{\delta}X_I} = 
- \partial_a{\pi}^{aI} \ . \end{equation}

\noindent Making use of this in the phase space action (in the first 
of the two forms given above) we obtain 

\begin{equation} S = \int \dot{X}_IP^I - e_{tI}P^I \ . \end{equation}

\noindent This is the action describing the kinematics of the vector 
$e_I$ that describes the deformations of spatial hypersurfaces in 
spacetime \cite{Dirac}. 

We can now trade the four vector $e_I$ for the lapse and shifts, just 
as we did above. This means that we write 

\begin{equation} e_{tI} = Nn_I + N^a\partial_aX_I \ , \end{equation} 

\noindent where the vector $n_I$, as defined in eq. (\ref{17}), obeys 

\begin{equation} n^I N^a\partial_aX_I = 0 \ . \end{equation}

\noindent We can also add any matter action (with metrical couplings) 
to the phase space action. In this way we arrive at 

\begin{eqnarray} S = \int \dot{X}_IP^I + N\frac{1}{3!}{\epsilon^{IJK}}_L
{\epsilon}^{abc}\partial_aX_I\partial_bX_J\partial_cX_K P^L
- N^a\partial_aX_IP^I + S_m \ . \end{eqnarray}

\noindent This is the action of a parametrized field theory on a 
flat background \cite{Dirac}; 
the constraint algebra is the same as that of general relativity. 

In conclusion, we have realised the equivalence between a sector of a
BF theory and Smolin's $G=0$ limit of Einstein's gravity, successfully
introduced matter terms in these models, and shown that they are equivalent
to parametrized field theory on a flat background provided taht spacetime
is simply connected.

\end{document}